\documentclass[aps,prl,twocolumn,groupedaddress]{revtex4}

\usepackage{graphicx}
\usepackage{dcolumn}
\usepackage{bm}

\begin{document}


\title{Observation of Vertical Betatron Sideband due to Electron Clouds in the KEKB LER}


\author{J. W. Flanagan}
\email[]{john.flanagan@kek.jp}
\affiliation{High Energy Accelerator Research Organization (KEK)}

\author{K. Ohmi}
\affiliation{High Energy Accelerator Research Organization (KEK)}

\author{H. Fukuma}
\affiliation{High Energy Accelerator Research Organization (KEK)}

\author{S. Hiramatsu}
\affiliation{High Energy Accelerator Research Organization (KEK)}

\author{M. Tobiyama}
\affiliation{High Energy Accelerator Research Organization (KEK)}

\author{E. Perevedentsev}
\affiliation{Budker Institute of Nuclear Physics}


\date{\today}

\begin{abstract}
The effects of electron clouds on positively-charged beams
have been an active area of research in recent years at
particle accelerators around the world.  Transverse beam-size
blow-up due to electron clouds has been observed in some
machines, and is considered to be a major limiting factor in the
development of higher-current, higher-luminosity electron-positron
colliders.  The leading proposed mechanism
for beam blow-up is the excitation of a fast head-tail instability
due to short-range wakes within the electron cloud.
We present here observations of betatron oscillation sidebands in
bunch-by-bunch spectra that may provide direct evidence of
such head-tail motion in a positron beam.
\end{abstract}


\maketitle

The development of clouds of electrons in positively-charged-beam storage
rings has been observed at several machines, including the KEKB Low Energy
Ring (LER),
a $3.5$ GeV positron storage ring which is part of the KEK
B-Factory.  Observations at the KEKB LER
of betatron tune shifts along a bunch train via gated tune meter
\cite{ref:ieiri_gatedtune}, and of transverse bunch size along the train via
high-speed gated camera \cite{ref:flanagan_gatedcamera} and streak camera
\cite{ref:streakcamera}, show a characteristic increase of transverse tune
shifts and beam size starting near the head of the train, reaching
saturation at some point along the train.  Simulations of electron
cloud density due to photo-electrons being drawn towards the positron
beam have shown a similar build up of cloud density along the train,
reaching saturation at some point \cite{ref:ohmi_1995, ref:zimmermann_cloud}.
Electrons from the cloud have also
been measured directly via electrode \cite{ref:ohnishi_rfa}.
Solenoids have been wound around approximately $95\%$ of the drift space in
the LER, with a maximum field at the center of the beam pipe of 45 Gauss
\cite{ref:fukuma_solenoids}.  The beam size blow-up has been observed to
occur above a threshold average bunch current of $\sim0.35$ mA/bunch
at 4-rf bucket spacing between bunches
with the solenoids off;  this
threshold is raised when the solenoids are powered on\cite{ref:fukuma_blowup}.
The beam blow-up has been found to reduce the specific luminosity of the
affected bunches 
\cite{ref:fukuma_blowup}.

One proposed mechanism for the beam blow-up due to the
presence of electron clouds is a strong head-tail instability caused
by wake fields
created
by the passage of the bunch particles through the
cloud \cite{ref:ohmi_zimmermann}.  Attempts have been made to observe
this head-tail motion directly via streak camera \cite{ref:streakcamera},
but have been unsuccessful, possibly due to a lack of sufficient light
intensity.  A vertical sideband peak has been reported for a proton
beam at the CERN SPS which could be an indication of head-tail motion
\cite{ref:arduini}, though no clear signature has yet been reported
at a positron machine.  We report here on observations of a sideband peak,
above the betatron tune, which may provide direct evidence of such a
coupled-mode spectral peak in a positron beam.

The sideband peak first appears near the bunch-current
threshold of beam blow-up -- the sidebands cannot be seen when the
average bunch current is below the beam-blow-up threshold,
and can be seen when the average bunch current
is over the threshold.  In addition, the presence of the sideband is
affected by the electron-cloud-suppression solenoids;  for example, it
has been observed to appear in bunches at $1$ mA per bunch and a 4-bucket
spacing only when the solenoids are turned off, and does not appear when the
solenoids are turned on.  These behaviors cannot be explained by ordinary
head-tail mechanisms;  we conclude that the sidebands, like the beam
blow-up, are caused by the presence of electron clouds.


\begin{figure}
\includegraphics[width=3.4in, height=2in]{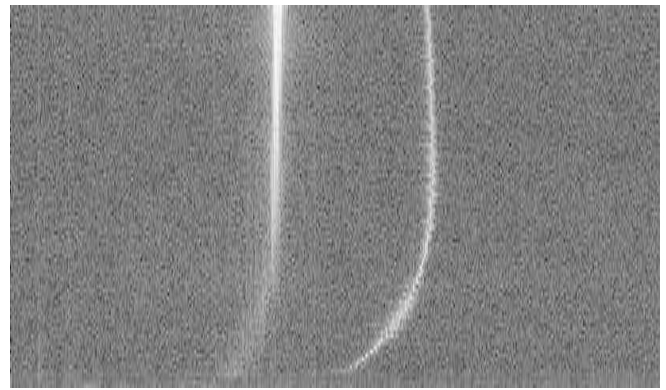}%
\caption{\label{fig:twodspecplot}Two-dimensional plot of vertical bunch
spectrum versus bunch number.  The horizontal axis is fractional tune,
from $0.5$ on the left edge to $0.7$ on the right edge.  The vertical axis
is bunch number in the train, from 1 on the bottom edge to 100 on the
top edge.  The bunches in the train are spaced 4 RF buckets
(about 8 ns) apart.  The bright, curved line on the left is the vertical
betatron tune, made visible by reducing the bunch-by-bunch feedback gain
by 6 dB from the level usually used for stable operation.  The line on the
right is the sideband.}
\end{figure}

Observations were made using signals taken from a pair of Beam 
Position Monitor (BPM) electrodes, which are mounted on the beam pipe,
and measure 6 mm in diameter.  The difference signal from two electrodes
on opposite sides of the beam pipe is detected at $2.032$ GHz 
(=$4 \times f_{RF}$), and recorded by the Bunch Oscillation Recorder (BOR),
which is a diagnostic tool in the bunch-by-bunch feedback system
\cite{ref:monitor_nim,ref:fb_prstab}.
The BOR itself consists of an 8-bit digitizer front-end, with a 20-MByte
memory, which is capable of storing one beam centroid position measurement per
bunch for all 5120 RF buckets in the ring over 4096 turns.

\begin{figure}
\includegraphics[width=3.4in]{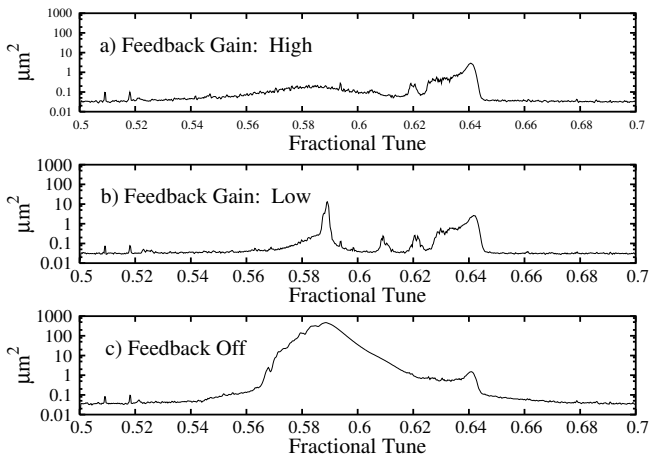}%
\caption{\label{fig:fbspectra}Averaged spectra of all bunches with the
feedback gain a) high, b) low and c) set to zero.
The vertical betatron peak is
visible at $0.588$, and the sideband peak can be seen around $0.64$.}
\end{figure}

Data were taken at the LER on 24 June 2004, in single-beam mode
(no colliding bunches in the HER) and
with the majority of the solenoids turned off.  The fill pattern consisted
of four trains of bunches, spaced evenly around the ring.  Each train
consisted of 100 bunches, spaced 4 RF buckets ($\approx 8$ ns) apart.
In Figure \ref{fig:twodspecplot}, the spectrum for each bunch is plotted, with
fractional tune on the horizontal axis and bunch number on the vertical
axis.  The betatron tune (made visible by lowering the gain of the
bunch-by-bunch feedback system) is seen as the left curved line.  The
betatron tune is seen to shift successively higher as one moves from
the head of the train towards the tail, saturating at around the 40th
bunch.  To the right of the $\nu_{y}$ peak is the sideband peak, which
likewise shifts along the train, until saturating a little after the
$\nu_{y}$ peak.
In experiments that have been performed over the past year, it has 
been observed that changing the vertical betatron tune causes the sideband
peak to shift by an equal amount, and in the same direction as that of
the betatron tune.  Changing the horizontal tune has no effect on
this sideband.

Figure \ref{fig:fbspectra}a shows the observed bunch spectra with the
feedback gain set at the nominal value for physics running, $-9.45$ dB.
In this plot, the Fourier power spectrum of each bunch in the ring is
calculated individually, then the power spectra of all bunches are averaged
together.  The horizontal scale is in units of
fractional tune, and the vertical scale is in units of ${\mu}m^{2}$.
The vertical betatron tune can be seen as a broad,
low peak at a fractional tune of approximately $0.58$-$0.59$.  To the
right of it can be seen the sideband peak at approximately $0.64$.
The broad, pedestal-like tail to the left of the peak is due to the
projection of a succession of narrow peaks, one for each bunch, which
have lower tunes near the head of the train.

The vertical gain of the feedback system was lowered by 6 dB, to the
point where the beam started to become slightly unstable, as seen in
oscilloscope traces and as reflected in a reduced lifetime for the beam.
Under these conditions, the vertical betatron peak becomes enhanced,
as shown in Fig. \ref{fig:fbspectra}b, however the sideband peak
amplitude is virtually unchanged.

Finally, the feedback was turned off entirely.  The BOR was set to record the
4096 turns immediately following the feedback being turned off.  As seen
in Fig. \ref{fig:fbspectra}c, the betatron
peak grows enormously, but the sideband peak height is again essentially
unchanged.  This indicates that this peak does not respond to dipole kicks
from the feedback system.  The estimated amplitude of the motion at this
peak is approximately $1.6$ $\mu$m,or half of one percent of the vertical
beam size of $320$ $\mu$m at the pickup location.


%

\begin{figure}
\includegraphics[width=3.4in]{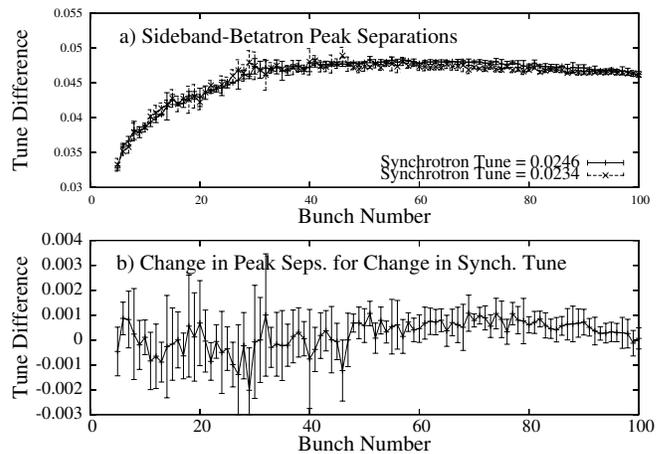}%
\caption{\label{fig:deltanus}Effect of changing synchrotron tune on
the separation between sideband peak and betatron peak.  In a), the
sideband-betatron peak separation is plotted along the bunch train for
$\nu_{s}=0.0246$ (solid lines) and $\nu_{s}=0.0234$ (dashed lines).
In b), the difference between the two curves is plotted.  Statistical
1-sigma error bars are shown.}
\end{figure}

Experiments were done with changing the synchrotron tune.  In one set of
measurements, the RF voltage was reduced, which lowered
the synchrotron tune by $0.0012$.  The position
of the sideband relative to the vertical betatron tune for both
values of $\nu_{s}$ are shown in Figure \ref{fig:deltanus}a;  the
sidebands are visible starting from the fifth bunch in the train.
The difference between the two curves is shown in Figure \ref{fig:deltanus}b.
The average of the peak separation over all bunches
is not statistically different from zero.

Observations have also been made using the same BOR memory recorder,
but using a fast photo-multiplier tube (PMT) as input device instead of a BPM
electrode.  A Hamamatsu H6780 PMT, was set up to record the light
intensity from a focused image of the beam using synchrotron radiation.
The image was partially obscured in the vertical direction, leaving 
only the upper edge of the image visible.  The spectra obtained via
PMT were identical to those obtained from the BPM electrode, though
with a lower signal-to-noise ratio.  The amplitude of
the peak seen by PMT can only be crudely estimated, but agrees roughly
with that seen by the BPM electrode.  One feature that the PMT can
detect that the BPM cannot is changes in the beam size.

\begin{figure}
\includegraphics[width=3.4in]{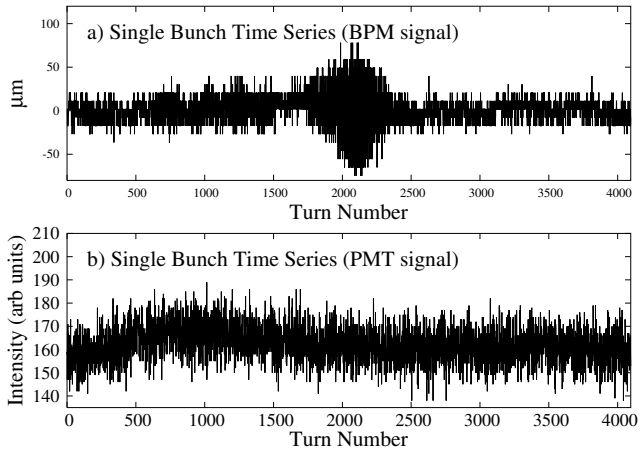}%
\caption{\label{fig:timeseries}Example time series of single bunches
taken via a) BPM and b) PMT.  Different bunches are shown for each
detector;  the data were taken within one minute of each other.
A burst-like behavior is visible in the BPM signal.  A fast ramp-up
behavior with a similar rise-time as the BPM burst is seen in
the PMT signal, followed by a gradual ramp-down.}
\end{figure}

The time-series data of the BPM, shown in Fig. \ref{fig:timeseries}a,
reveal a burst-like time structure to the sideband oscillations.
The sideband peak is present as a low level oscillation that suddenly
grows and damps in a burst lasting $\sim500$ turns (5 ms).  A break down
of the data into 512-turn slices shows that the sideband peak is seen most
strongly during the burst, and disappears entirely just after the burst.

In the PMT data, as seen in Fig. \ref{fig:timeseries}b,
a similar 500-turn-duration phenomenon is observable
wherein the light level (beam size) increases over the course of 500 turns,
then slowly damps afterwards, over the course of $\sim1500$ turns.  A
slice-by-slice breakdown of such events reveals that the sideband peak
is a maximum during the ramp-up, and disappears momentarily just after
the burst.

The two sets of observations suggest that in the
blown-up state, a series of bursts and quiescent periods alternate.  During
the bursts of violent dipole motion, the beam size increases by a
further $\approx5\%$ from its already blown-up state.  After it blows up,
the dipole motion is temporarily absent, as the emittance of the beam
damps down.

\begin{figure}
\includegraphics[width=2.4in]{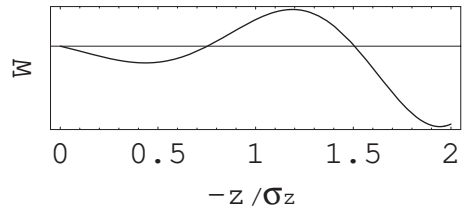}%
\caption{\label{fig:modelwake}Model focusing wake.  The horizontal axis
is longitudinal position normalized to the bunch length.}
\end{figure}
\begin{figure}
\includegraphics[width=3.4in]{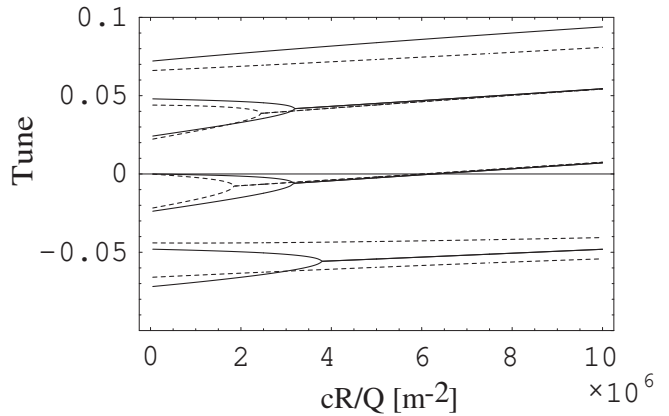}%
\caption{\label{fig:airns}Example mode spectrum for model focusing
wake at $\nu_{s}=0.022$ (dashed lines) and $\nu_{s}=0.024$ (solid lines).}
\end{figure}
\begin{figure}
\includegraphics[width=3.4in]{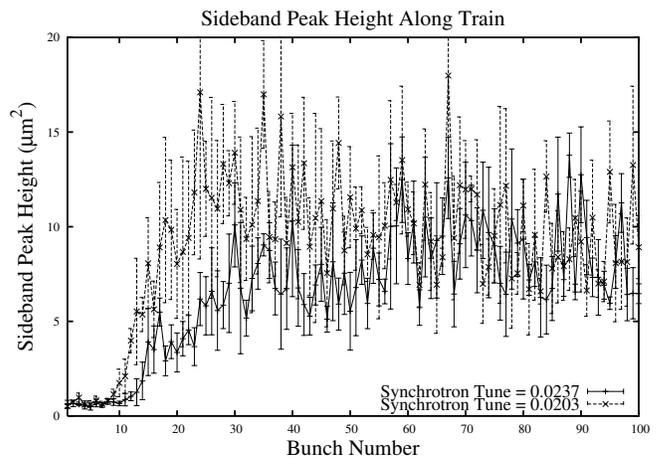}%
\caption{\label{fig:peakheight_nus}Peak heights of maximum-amplitude 
frequency bins in sideband region of bunch spectrum, plotted for the 
first 100 bunches in the train, for two values of $\nu_{s}$.  Error 
bars are 1-sigma statistical error bars.}

\end{figure}

One possible interpretation for this sideband is that it is a signature
of mode-coupling due to the head-tail instability predicted by Ohmi, 
Zimmermann and Perevedentsev \cite{ref:ohmi_zimmermann_perevedentsev}.
A notable feature of the side band is that it occurs on the upper side
of the betatron peak, which suggests that the effective wake function in
the region of the tail of the bunch is a focusing wake.  A possible
mechanism for producing such a wake is a pinching effect on the electron
cloud.  Simulations of wakes that change from defocusing to
focusing with distance along the bunch have been found in simulations
using the KEKB parameters
\cite{ref:ohmi_zimmermann_perevedentsev, ref:rumolo_zimmermann}.

When the synchrotron tune is changed, the average separation between the 
sideband peak and the betatron peak does not change significantly.  In the case
of strong head-tail instability, the coupled mode frequency does not
necessarily depend strongly on $\nu_{s}$.  As an illustration, mode
spectra were generated using a toy model
with an airbag charge distribution and a simple effective wake,
shown in Fig. \ref{fig:modelwake}, which uses a resonator-like wake $W$,
increasing along $(-z)$ to represent the enhancement of the wake near the
tail of the bunch due to pinching of the electron cloud:

\begin{equation}
W(z)=c\frac{R}{Q}e^{-\alpha z/c}\sin{\omega_{R} \frac{z}{c}},
\end{equation}
where $\alpha=\omega_{R}/4$, and $\omega_R=2\pi\times40$ GHz.
(Note: the oscillation frequency of cloud electrons as calculated from
the LER beam size and positron charge density is $\sim 2\pi \times 43$ GHz.)

Plots of mode spectra as a function of effective $R/Q$ are
shown in Fig. \ref{fig:airns} for synchrotron tunes of $0.022$ and $0.024$.
As can be seen, the tune of the coupled mode in the region
far above the coupling threshold does not change significantly
with the synchrotron tune.
However, the coupling point between the $l=+1$ mode ($\nu_{y}+\nu_{s}$)
and the $l=+2$ mode ($\nu_{y}+2\nu_{s}$) shifts to the right.
Since the electron cloud density increases along the leading bunches 
of the train, this change in the threshold would lead one to expect the
position of the first bunch to exhibit the sideband should shift as
the synchrotron tune is changed.  To investigate this behavior near
the threshold, data originally taken on 23 December 2003
at two different values
of the synchrotron tune were re-examined.  The LER was in single beam mode,
with all solenoids off.  The bunches were stored at a four-bucket spacing,
at a bunch current of $0.52$ mA/bunch.  Due to the lower bunch current,
the growth of the sideband peak is more gradual in this data set than
it is in the July 2004 data set.  (Due to a high feedback gain,
the betatron peak is not pronounced enough to measure.)  One set, of
four measurements, was taken at an RF voltage of 8 MV, and the other
set, of three measurements, was taken at 6 MV.
The synchrotron tunes of the two
sets, as measured from the synchrotron peak visible in the spectra, were
$0.0237$ and $0.0203$, respectively.  The maximum-height frequency bin
in the region of the sideband was found for each bunch in the train, and
the peak heights of those maximum bins were averaged together within each
set.  The average peak heights, and 1-sigma statistical error bars at
each synchrotron tune are plotted as a function of bunch number along the
train in Fig. \ref{fig:peakheight_nus}.  As can be seen, the development
of the sideband peak height occurs earlier in the train at the lower
synchrotron tune, in agreement with expectation.


Simulations and linear theory also indicate that for a given cloud
density, larger beam sizes should be more stable due to a weaker
beam-cloud interaction \cite{ref:ohmi_zimmermann_perevedentsev}.
The burst-like activity of the blown-up beam
may be the result of the beam size varying around some threshold value,
as the bunch alternates between states of emittance growth due to the
instability occurring when the beam size is below the threshold, and
the beam size damping down once over the threshold.

A betatron sideband peak has been found in the vertical tune spectrum
of positron bunches in the presence of beam-size blow-up due to
electron clouds.  The sideband peak is on the upper side of the betatron 
peak in terms of fractional
tune, first appears early in the bunch train, and the separation between
this peak and the betatron tune peak increases going along the train, until
it saturates at a certain point.  The best explanation for it is that
it is a signature of the head-tail instability hypothesized to explain
transverse beam blow-up due to electron clouds.  The presence of this
sideband peak also provides a sensitive diagnostic for the presence of
electron clouds.

The authors would like to thank Professor K. Oide for his support of this
work, and Drs. Y. Funakoshi, T. Ieiri, H. Ikeda, H. Koiso, M. Masuzawa
and A. Valishev for many fruitful discussions.

\end{document}